\documentclass[12pt]{iopart}

\usepackage{iopams}

\def\vps{{\vec\psi}}
\def\vph{{\vec \phi}}
\def\rhb{\mbox{\boldmath{$\rho$}}}
\def\bs{\breve\sigma}
\def\tr{{\rm tr}}
\def\im{{\rm Imp}}
\def\intx{\int\limits_{-\infty}^{+\infty} {\rm d}x}
\def\into{\int\limits_0^{+\infty}\frac{{\rm d}\omega}{2\pi}}
\def\sg{{\rm sg}}

\begin{document} 
   
\title{Impurity and quaternions in nonrelativistic scattering from a quantum memory}

\author{D Margetis$^{1,2}$ and M G Grillakis$^1$}

\address{${}^1$ Department of Mathematics, University of Maryland, College Park, MD 20742, USA}
\address{${}^2$ Institute for Physical Science and Technology, University of Maryland,
College Park, MD 20742, USA}

\ead{\mailto{dio@math.umd.edu} and \mailto{mng@math.umd.edu}}

\begin{abstract}
Models of quantum computing rely on transformations of the states of a quantum memory.
We study mathematical aspects of a model proposed by Wu in which the memory state is changed via 
the scattering of incoming particles.
This operation causes the memory content to deviate from a pure state, i.e. induces impurity.
For nonrelativistic particles scattered from a two-state memory and sufficiently
general interaction potentials in 1+1 dimensions, we express impurity 
in terms of quaternionic commutators. In this context, pure memory states correspond to 
null hyperbolic quaternions. In the case with point interactions,
the scattering process amounts to appropriate
rotations of quaternions in the frequency domain. Our work complements 
a previous analysis by Margetis and Myers (2006 {\it J.\ Phys.\ A} {\bf 39} 11567--11581).\looseness=-1
\end{abstract}

\vspace{2pc}

\submitto{J. Phys. A: Math. Theor.}
\maketitle 

\section{Introduction}

Ideally, quantum computations are performed via transforming pure states of a 
physical system called `quantum memory' to 
other pure states; see e.g.~\cite{benioff80,albert83,feynman85,deutch85,shor95,jaeger06}. 
In this context, memory states transform unitarily. 
In most systems pure states may degrade to mixed states. This phenomenon amounts to
decoherence. A well-known kind of decoherence is caused by extraneous influences
unrelated to memory operations~\cite{jaeger06,heiss02}. 

Recently, Wu~\cite{wu02,wu03,wu05} introduced spatial variables in quantum computations by
viewing the quantum memory as a scatterer: incoming
particles are scattered from the memory and change its content.
In this setting, unitary transforms apply to the combined system of memory and particles.
The memory states do not transform unitarily unless the incoming 
signal is `admissible'. In one space dimension, single-frequency waves are admissible~\cite{wu03}.
In practice, however, incoming signals are pulses of finite duration. Thus, their use 
leads to additional decoherence, which we term `impurity'.
This kind of decoherence is connected specifically to memory operations, as was first 
discussed in~\cite{wu03}. 
The impurity of a two-state memory was analyzed via the
relativistic~\cite{wu03} and nonrelativistic~\cite{margetismyers} Schr\"odinger equations.
In~\cite{margetismyers} the memory is allowed to interact with incoming particles
only at one point by use of the pseudo-potential derived in~\cite{wu02}.\looseness=-1

In the present paper we extend the nonrelativistic formulation 
of~\cite{margetismyers} to reasonably general interaction potentials in one space dimension.
Our starting point is to model the interaction potential as an imaginary quaternion~\cite{hamilton,weyl,naber,ward}. 
In this formalism, the impurity measure of~\cite{margetismyers} is expressed naturally in terms of an appropriate  
norm that depends on quaternionic commutators; see proposition~I of section~\ref{subsec:imp}. 
In this context, pure states correspond to null hyperbolic quaternions~\cite{mcfarlane}.
For point interactions, scattering from the memory amounts to appropriate rotations of quaternions
in the frequency domain. This approach offers additional insight into properties of the impurity 
measure used in~\cite{margetismyers}, and is amenable to computations.\looseness=-1 

The paper is organized as follows. In section~\ref{sec:formulation} we formulate
the problem of impurity for a two-state quantum memory as a scattering problem
with two coupled channels and general interaction potential in one space dimension:
in section~\ref{subsec:eq-mot} we formulate the equations of motion by treating
a local interaction potential as a quaternion; in section~\ref{subsec:imp} we express
the time evolution of the impurity measure used in~\cite{margetismyers} in terms of commutators of quaternions;
and in section~\ref{subsec:ext} we describe an extension of this formulation to nonlocal
interaction potentials.
In section~\ref{sec:soln} we describe the general solution by invoking
discrete schemes for amplitudes of suitable Fourier transforms in time.
In section~\ref{sec:point-inter} we revisit the case with point interactions by use of the present formalism:
in section~\ref{subsec:even} we focus on even wavefunctions; and in section~\ref{subsec:odd}
we treat odd wavefunctions. In section~\ref{sec:conclusion} we summarize our results
and discuss open problems. 
Throughout the analysis we apply units with $\hbar^2/(2m)=1$ where $m$ is the particle mass; and denote
quaternions by boldface symbols distinct from vectors.

\section{Formulation}
\label{sec:formulation}

In this section we describe the equations of motion for a general interaction potential.
Subsequently, we derive an explicit formula for the impurity used in~\cite{margetismyers}.\looseness=-1

\subsection{Equations of motion}
\label{subsec:eq-mot}

For a two-state quantum memory~\cite{wu02},
the field of the particle-memory system is the $2\times 1$ (column) vector\looseness=-1
\begin{equation}
\vps(x,t)= \left[\begin{array}{c} \psi_1(x,t) \\
\psi_2(x,t) \end{array}\right]\qquad -\infty< x,\ t< +\infty~, \label{eq:Psi}
\end{equation}where $\psi_j(x,t)$ ($j=1,\,2$) are scalar, square integrable 
functions in 1+1 dimensions. The vector field $\vps$ solves the Schr\"odinger equation,
\begin{equation}
i\partial_{t}\vps=-\partial^{2}_{x}\vps - i{\bf q}(x)\vps~,\label{eq:S_Eq}
\end{equation}
where $-i{\bf q}(x)$ represents the interaction potential and ${\bf q}$
is a $2\times 2$ skew-adjoint matrix; see~(\ref{eq:qstar}) below.

The ${\bf q}$ of~(\ref{eq:S_Eq}) is written as 
\begin{equation}
{\bf q}(x)=\sum_a g^{a}(x)\bs_{a}~,\quad a=1,\,2,\,3~,\label{eq:q-def}
\end{equation}
which we call an `imaginary quaternion'~\cite{hamilton};
$g^a(x)$ are given real functions, $\bs_a:=i\sigma_a$, and 
$\sigma_\mu$ ($\mu=0,\,1,\,2,\,3$) are the Pauli matrices with the usual convention $\sigma_{0}={\bf 1}$,
\begin{equation}
\sigma_{1}=\left[\matrix{0&  1\cr 1&  0\cr}\right]\ ,\ 
\sigma_{2}=\left[\matrix{0& -i\cr i&  0\cr}\right]\ ,\
\sigma_{3}=\left[\matrix{1&  0\cr 0&  -1\cr}\right]~. \label{eq:sigma}
\end{equation}
So, $\bs_{a}$ have the following properties.
\begin{eqnarray}
\lefteqn{\bs^{\ \dag}_{a}=-\bs_{a}~,\quad \bs^{2}_{a}=-{\bf 1}~,} \nonumber\\
&&\bs_{a}\bs_{b}=-\sum_c \epsilon_{abc}\,\bs_{c}~;
\quad a\not= b~,\quad  a,b,c=1,2,3~,\label{eq:sigma-prop}
\end{eqnarray}
where $\sigma^\dag$ denotes the conjugate transpose (Hermitian conjugate) of $\sigma$, and
$\epsilon_{abc}$ is the Levi-Civita symbol. Hence, we ensure that
\begin{equation}
{\bf q}^{\ \dag}(x)=-{\bf q}(x)~.\label{eq:qstar}
\end{equation}
A simplified form of~(\ref{eq:S_Eq}) is
\begin{equation}
i\partial_{t}\vps=-\partial^{2}_{x}\vps -
i\big[g_{e}(x){\bf u}_{e}+g_{o}(x){\bf u}_{o}\big]\vps~, \label{eq:S_Eq-odev}
\end{equation}
where ${\bf u}_{e,o}$ are unit-length imaginary quaternions and $g_{e,o}(x)$ are
even and odd functions, respectively. 

We note in passing that the `discrete version' of~(\ref{eq:S_Eq}) reads
\begin{equation}
i\partial_{t}\vps=-\partial^{2}_{x}\vps -
\sum_j i{\bf q}_{j}\delta(x-x_{j}){\vec\psi}~,\label{eq:discr}
\end{equation}
where $\{{\bf q}_{j}={\bf q}(x_j)\}$ ($j$: integer) is a sequence of imaginary quaternions.
By convolution of~(\ref{eq:discr}) with an appropriate kernel and for
sufficiently dense partition $\{ x_j\}$, the resulting solution
can be arbitrarily close to the solution of~(\ref{eq:S_Eq}).
Because~(\ref{eq:discr}) is amenable to numerical computations, we discuss 
the relevant solutions in detail in section~\ref{sec:soln}.

For later convenience, we introduce the Fourier transform in time of $\vps(x,t)$ by
assuming that this signal contains only positive frequencies. With the definition
\begin{equation}
\vps(x,t)=\into\ e^{-i\omega t}\,\vph(x,\omega)~,
\label{eq:FT-psi}
\end{equation}
the equation of motion~(\ref{eq:S_Eq}) transforms to
\begin{equation}
\omega \vph(x,\omega)=-\partial_x^2\vph(x,\omega)-i{\bf q}(x)\vph(x,\omega)~.
\label{eq:S_Eq-FT}
\end{equation}

\subsection{Impurity}
\label{subsec:imp}

Next, we analyze impurity on the basis of~(\ref{eq:S_Eq}). 
The reduced density matrix $\rhb(t)$ for the memory is the $2\times 2$ matrix~\cite{margetismyers}
\begin{equation} 
\rhb(t) := \intx\ \vps(x,t)\vps^{\ \dag}(x,t)~,
\label{eq:rho-def}
\end{equation} 
which is obtained by tracing out the spatial variables. Of particular interest is the limit
\begin{equation}
{\bf M}:= \lim_{t\to +\infty}\rhb(t)~,\label{eq:M-def}
\end{equation}
which is connected to the final memory state.
For vanishing impurity, ${\bf M}^2={\bf M}$ and $\tr({\bf M}^2)=1$. 
In~\cite{margetismyers}, the impurity measure is defined by
\begin{equation}
\im({\bf M}) := [1 - \tr({\bf M}^2)]^{1/2}~.\label{eq:im}
\end{equation} 

In this subsection, we describe the ${\bf M}$ of~(\ref{eq:M-def}). 
For this purpose, we form the density matrix 
\begin{eqnarray}
{\bf Q}(x,t)&:=&\vps\vps^{\ \dag}=\left[\matrix{\psi_{1}\psi_{1}^{\ast}&\psi_{1}\psi^{\ast}_{2}\cr
\psi_{2}\psi^{\ast}_{1}&\psi_{2}\psi^{\ast}_{2}\cr}\right]\nonumber\\
{} &=&-i\biggl(ir^0\sigma_0+\sum_a r^a\bs_a\biggr)
=:-i{\bf r}(x,t),\label{eq:Q-def}
\end{eqnarray}
where $\psi^{\ast}_j$ is the complex conjugate of $\psi_j$,
and the coefficients $r^{\mu}$ ($\mu =0,1,2,3$) are 
\begin{eqnarray}
r^{0}&=&
\frac{1}{2}\big(\vert\psi_{1}\vert^{2}+\vert\psi_{2}\vert^{2}\big)~,\qquad
r^{1}=
\frac{1}{2}\big(\psi_{1}\psi^{\ast}_{2}+\psi^{\ast}_{1}\psi_{2}\big)~,\nonumber\\
r^{2}&=&
\frac{1}{2i}\big(\psi^{\ast}_{1}\psi_{2}-\psi_{1}\psi^{\ast}_{2}\big)~,\qquad
r^{3}=\frac{1}{2}\big(\vert\psi_{1}\vert^2-\vert\psi^{\ast}_{2}\vert^2\big)~.\label{eq:rmu-def}
\end{eqnarray}
It follows that
\begin{equation}
\rhb(t)=-i\intx\,{\bf r}(x,t)=:-i{\bf m}(t)~,\quad {\bf m}(t):=\intx\,{\bf r}(x,t)~,
\label{eq:m-def}
\end{equation}
\begin{equation}
{\bf M}=-i\lim_{t\to +\infty}{\bf m}(t)=:-i{\bf m}_{out}~.\label{eq:mout-M}
\end{equation}

An important remark is in order. With ${\bf r}=ir^0\sigma_0+\sum_a r^a\bs_a$ by~(\ref{eq:Q-def}), 
the dual of ${\bf r}$ is defined by
\begin{equation}
{\bf r}^{\rm d}:=ir^{0}\sigma_{0}-\sum_a r^{a}\bs_{a}~.\label{eq:dual}
\end{equation}
By virtue of~(\ref{eq:rmu-def}), we have 
\begin{equation}
{\bf r}{\bf r}^{\rm d}=:\langle {\bf r}\,,\,{\bf r}\rangle \equiv 0~,\label{eq:null}
\end{equation}
i.e., ${\bf r}$ is identified with a null hyperbolic
quaternion~\cite{mcfarlane}; $\langle\cdot\,,\,\cdot\rangle$ is 
the Minkowski inner product by identifying ${\bf r}$ with the four-vector
($r^0\,,r^1,\,r^2,\,r^3$). Note that the ${\bf m}$ in~(\ref{eq:m-def})
is an integral of null quaternions. Because of the convexity of the characteristic
cone~\cite{naber}, this integral produces either a time-like (${\bf m}{\bf m}^{\rm d}<0$) or a
null (${\bf m}{\bf m}^{\rm d}=0$) 
quaternion. This property ensures that the $1-\tr({\bf M}^2)$ in~(\ref{eq:im})
is non-negative; see appendix.

We now derive an equation of motion for ${\bf Q}(x,t)$. With the system Hamiltonian
${\bf H}:=-{\bf 1}\partial^{2}_{x}-i{\bf q}(x)$, (\ref{eq:S_Eq}) becomes
$i\partial_t\vps={\bf H}\vps$. By~(\ref{eq:Q-def}), we readily obtain
\begin{equation}
\partial_{t}{\bf Q}=-i\big({\bf H}_{\rightarrow}{\vps}\big){\vps^{\ \dag}}
+i{\vps}\big({\vps}^{\ \dag}_{\leftarrow}{\bf H}\big)~,\label{eq:Q-mot}
\end{equation}where $\vps^{\ \dag}_{\leftarrow} {\bf H}$ denotes the action of ${\bf H}$
on $\vps^{\ \dag}$ from the left. Equation~(\ref{eq:Q-mot}) is recast to the conservation law 
\begin{equation}
\partial_{t}{\bf Q}+\partial_{x}{\bf P} =\big[{\bf Q}(x,t)\,,\,{\bf q}(x,t)\big]~,\label{eq:cons}
\end{equation}
where $[{\bf Q}\,,\, {\bf q}]:={\bf Qq}-{\bf qQ}$ is the commutator of two quaternions and
${\bf P}$ is the `flux matrix'  
\begin{equation}
{\bf P}={1\over i}\left[
\big(\partial_{x}{\vec\psi}\big){\vps}^{\ \dag}-\vps\big(\partial_x\vps^{\ \dag}\big)\right]~.\label{eq:P-def}
\end{equation} 

By~(\ref{eq:m-def}) and~(\ref{eq:mout-M}), we need to integrate~(\ref{eq:Q-mot}) over space and then time. 
Accordingly, we obtain
\begin{equation}
\Delta {\bf m}:={\bf m}_{out}-{\bf m}_{in}=
\int\limits_{-\infty}^{+\infty}\int\limits_{-\infty}^{+\infty}{\rm d}t\,{\rm d}x\,
\left[{\bf r}(x,t)\, ,\, {\bf q}(x)\right]~,\label{eq:Deltam}
\end{equation}
where ${\bf m}_{in}:=\lim_{t\to -\infty}{\bf m}(t)$ is given.
Consequently, 
\begin{equation}
{\bf M}=-i({\bf m}_{in}+\Delta{\bf m})~.\label{eq:M-moutin}
\end{equation}

It is convenient to rewrite the $\Delta{\bf m}$ of~(\ref{eq:Deltam})
in terms of the appropriate Fourier transform. 
Plancherel's formula~\cite{dymmckean} and definition~(\ref{eq:Q-def}) give
\begin{equation}
\int\limits_{-\infty}^{+\infty}{\rm d}t\ {\bf r}(x,t) =
\into\ \widetilde{\bf r}(x,\omega)~,\label{eq:Planch}
\end{equation}
where $\widetilde{\bf r}(x,\omega)$ is the quaternion corresponding to the frequency-domain density matrix
$\vec{\phi}(x,\omega){\vph}^{\ \dag}(x,\omega)$ by~(\ref{eq:FT-psi}), i.e.,
\begin{equation}
\widetilde{\bf r}(x,\omega):= i \vph(x,\omega)\vph^{\ \dag}(x,\omega)~,
\label{eq:hat-r}
\end{equation}
assuming that the signals have only positive-frequency content. Thus, we have the formula
\begin{equation}
\Delta{\bf m}=
\into\int\limits_{-\infty}^{+\infty}{\rm d}x\ 
\left[\widetilde{\bf r}(x,\omega)\,,\,{\bf q}(x)\right]~.\label{eq:Deltam-FT}
\end{equation}
We note in passing that~(\ref{eq:Deltam-FT}) can be formally generalized for higher space dimensions, where
$\vph(x,\omega)$ satisfies the vector Helmholtz equation.\looseness=-1

We now restrict attention to pure initial memory states.
The incoming vector field has the product form
\begin{equation}
\vps_{in}\sim \psi_{in}(x,t){\vec s}_{in}\qquad t\to -\infty~,\label{eq:inc}
\end{equation}
where $\psi_{in}(x,t)$ is the incoming particle wavefunction and ${\vec s}_{in}$
is the initial memory state. For example, we have~\cite{margetismyers}
\begin{equation} 
\psi_{in}(x,t)=\into\ e^{-i\omega t-i\sqrt{\omega}|x|}\,f(\omega)\cdot\left\{\begin{array}{lr}
1,& \mbox{even wave}\\
\sg(x),& \mbox{odd wave}\end{array}\right.~,\label{eq:psi-in-f}
\end{equation}
where $\sg(x)$ is the usual sign function, i.e., $\sg(x)=1$ if $x>0$, $\sg(x)=-1$ if $x<0$ and $\sg(0)=0$. 
The incoming quaternion ${\bf m}_{in}$ is~\cite{margetismyers}
\begin{equation}
{\bf m}_{in}=4\into\ \vert f(\omega)\vert^{2}\sqrt{\omega}\,
\big({\vec s}_{in}{\vec s}^{\ \dag}_{in}\big)~.\label{eq:min-pure}
\end{equation}
We view~(\ref{eq:min-pure}) as the general definition of $f(\omega)$ (without specifying any symmetry in $x$).
By analogy with~(\ref{eq:null}), ${\bf m}_{in}$ satisfies
\begin{equation}
{\bf m}_{in}{\bf m}_{in}^d=0~.\label{eq:m-null}
\end{equation}
Similarly, the condition $\tr({\bf M}^2)=1$ (pure final memory state) 
corresponds to ${\bf m}_{out}{\bf m}_{out}^d=0$.
So, pure states are described by null hyperbolic quaternions. This mapping is one to one, as
stated in proposition~I below.

The frequency profile $\vert f(\omega)\vert^{2}$ in~(\ref{eq:min-pure}) is chosen
so that ${\bf m}_{out}$ is as close to null as possible. We state the following proposition.
\vskip10pt

\noindent{\bf Proposition I.} {\it The impurity measure~(\ref{eq:im}) reads}
\begin{equation}
\im({\bf M})=\sqrt{2\vert\langle {\bf m}_{out}\,,\,{\bf m}_{out}\rangle\vert}=
\sqrt{2\vert2\langle{\bf m}_{in}\,,\,\Delta{\bf m}\rangle
+\langle\Delta{\bf m}\,,\,\Delta{\bf m}\rangle\vert}~,
\label{eq:prop}
\end{equation}
{\it where} $\langle\cdot\,,\,\cdot\rangle$ 
{\it is the Minkowski inner product defined in~(\ref{eq:null}).
(Thus, minimizing impurity is equivalent to minimizing the Minkowski norm corresponding
to} ${\bf m}_{out}$.) {\it In addition, zero impurity is equivalent
to} ${\bf m}_{out}$ {\it being a null hyperbolic quaternion.} 
\looseness=-1
\vskip10pt

A proof of~(\ref{eq:prop}) follows directly from the definition of ${\bf m}_{out}=i{\bf M}$ and~(\ref{eq:im}).
We sketch the main steps here; for details see~appendix. By the representation
of ${\bf m}_{out}$ in terms of the Pauli matrices we find
$\vert\langle {\bf m}_{out}\,,\,{\bf m}_{out}\rangle\vert = {\rm det}{\bf M}$.
Recall that the impurity measure is $\im({\bf M})=\sqrt{2\,({\rm det}{\bf M})}$~\cite{margetismyers}, and
for a pure initial state we have $\langle{\bf m}_{in}\,,\,{\bf m}_{in}\rangle=0$.

Proposition~I shows that definition~(\ref{eq:im}) for $\im({\bf M})$
is a natural choice: the deviation from a pure state is expressed in terms
of the `length' of a hyperbolic quaternion.
The impurity can be obtained from~(\ref{eq:prop}) combined with~(\ref{eq:Deltam-FT}), 
(\ref{eq:hat-r}) and~(\ref{eq:min-pure}) once $\vph$ is known. The reader is referred to section~\ref{sec:soln}
for details on~$\vph(x,\omega)$.

\subsection{Extension}
\label{subsec:ext}

In the case with a nonlocal interaction potential~\cite{wu02}, the equation of motion is
\begin{equation}
i\partial_{t}\vps=-\partial_{x}^{2}\vec{\psi} -\int\limits_{-\infty}^{+\infty} {\rm d}y\  
i{\bf q}(x,y)\vec{\psi}(y,t)~.\label{eq:S_Eq-nlc}
\end{equation}
The corresponding equation for ${\bf Q}=\vps{\vps}^{\ \dag}$ is
\begin{equation}
\partial_{t}{\bf Q} +\partial_{x}{\bf P} =\int\limits_{-\infty}^{+\infty}{\rm d}y\, \left\{
\vps(x,t)\vps^{\ \dag}(y,t){\bf q}(x,y)-{\bf q}(x,y)\vps(y,t)\vps^{\ \dag}(x,t)\right\}.
\label{eq:Q-mot-nlc}
\end{equation}

If ${\bf q}(x,y)={\bf q}V_\alpha(x)V_\beta(y)$ and $V_{\alpha,\beta}$
are scalar functions~\cite{wu02}, integration of~(\ref{eq:Q-mot-nlc}) yields
\begin{equation}
\partial_{t}{\bf m} =\big[{\bf r}_{V}\,,\,{\bf q}\big]~,
\label{eq:m-nonlc}
\end{equation}
where the quaternion ${\bf r}_{V}$ is defined by
\begin{equation}
{\bf r}_{V}= \int\limits_{-\infty}^{+\infty} {\rm d}y\ V_\beta(y)\vps(y,t)\int\limits_{-\infty}^{+\infty}{\rm d}x\  
V_{\alpha}(x)\vps^{\ \dag}(x,t)~.\label{eq:rV-def}
\end{equation}

\section{General solution scheme}
\label{sec:soln}

In this section we describe the Fourier transform $\vph(x,\omega)$ of $\vps(x,t)$ by solving~(\ref{eq:S_Eq-FT}).
We start with~(\ref{eq:discr}), the discrete analogue of~(\ref{eq:S_Eq}).
The Fourier decomposition~(\ref{eq:FT-psi}) reduces~(\ref{eq:discr}) to the form 
\begin{equation}
\omega\vph(x,\omega) =-\partial^{2}_{x}\vph(\omega ,x) -
\sum_j i\delta(x-x_j){\bf q}_{j}\,\vph_{j}~,\label{eq:vphi-discr}
\end{equation}
where $\vph_{j}:=\vph(x_j,\omega)$. 
The general solution of~(\ref{eq:vphi-discr}) is
\begin{equation}
\vph(x,\omega)=
{\vec a}_{+}(\omega)e^{i\sqrt{\omega}\ x}+{\vec a}_{-}(\omega)e^{-i\sqrt{\omega}\ x}+
\sum_j i{\bf q}_{j}{\vec\phi}_{j}G(x-x_{j})~,\label{eq:vph-soln-discr}
\end{equation}
where ${\vec a}_\pm$ are reasonably arbitrary vectors and $G(x,\omega)$ is the Green's function defined by
\begin{equation}
G(x,\omega)={i\over 4\sqrt{\omega}}
\biggl(e^{i\sqrt{\omega}|x|}-e^{-i\sqrt{\omega}|x|}\biggr)~.
\label{eq:G-def}
\end{equation}
By inspection of~(\ref{eq:vph-soln-discr}) and~(\ref{eq:G-def}), $\vph$ reads
\begin{equation}
\vph(x,\omega)={\vec d}_{+}(x,\omega)\,e^{i\sqrt{\omega}x}
+{\vec d}_{-}(x,\omega)\,e^{-i\sqrt{\omega}x}~,\label{eq:vph-alt}
\end{equation}where only the values ${\vec d}_\pm(x_j,\omega)$ matter.

In principle, the coefficients ${\vec d}_\pm$ can be determined from the incoming wavefunction $\vps_{in}$,
formula~(\ref{eq:inc}). In the limit $t\to -\infty$ we have~\cite{margetismyers}
\begin{equation}
\into\ {\vec d}_\pm(x,\omega)\,e^{\pm i\sqrt{\omega}x -i\omega t}\sim \psi_{in}(x,t)\,{\vec s}_{in}\qquad t\to -\infty~,
\label{eq:d-asymp}
\end{equation}where the upper (lower) sign is taken for $x<0$ ($x>0$). In particular, 
we set $x\to \mp\infty$~\cite{margetismyers}.
So, the last relation gives ${\vec d}_+(x,\cdot)$ if $x< -M_+$ and
${\vec d}_-(x,\cdot)$ if $x> M_-$ for sufficiently large $M_\pm$ by use of
the Fourier transform of $\psi_{in}$, e.g.~(\ref{eq:psi-in-f}). In the following,
we find $\vph(x,\cdot)$ everywhere via a scheme that determines ${\vec d}_\pm(x_j,\omega)$.

Next, we derive equations for ${\vec d}_{\pm,j}:={\vec d}_{\pm}(x_j,\omega)$ on the
basis of~(\ref{eq:vphi-discr}).
For this purpose, we introduce the four-component vectors
\begin{equation}
{\vec d}(x,\omega):=
\left[\matrix{{\vec d}_{+}\cr {\vec d}_{-}\cr}\right]~,
\qquad {\vec a}(\omega):=
\left[\matrix{{\vec a}_{+}\cr {\vec a}_{-}\cr}\right]~.
\label{eq:4-vect-da}
\end{equation}
Equation~(\ref{eq:vph-soln-discr}) reads
\begin{equation}
{\vec d}_{j}={\vec a}-\sum_{k}\sg(j-k)\,\Omega_{k}{\vec d}_{k}~,
\label{eq:alg-d}
\end{equation}
where ${\vec d}_j:={\vec d}(x_j,\omega)$ and $\Omega_j$ is the $4\times 4$ matrix
\begin{equation}
\Omega_{j}:=
{1\over 4\sqrt{\omega}}\left[\matrix{{\bf q}_{j}&w_{j}{\bf q}_{j}\cr 
-w^{\ast}_{j}{\bf q}_{j}& -{\bf q}_{j}\cr}\right]~,\qquad w_j:=e^{-2i\sqrt{\omega}\ x}~.\label{eq:Omega-def}
\end{equation}
By taking differences in~(\ref{eq:alg-d}) we find the equation
\begin{equation}
(I+\Omega_{j+1}){\vec d}_{j+1}=(I-\Omega_j){\vec d}_j~,\label{eq:alg-d-diff}
\end{equation}
where $I$ is the $4\times 4$ unit matrix. By the identity $(\Omega_j)^2\equiv 0$ we write~(\ref{eq:alg-d-diff}) as
\begin{equation}
{\vec d}_{j+1}=(I+\Omega_{j+1})^{-1}\,(I-\Omega_j){\vec d}_j=(I-\Omega_{j+1})\,(I-\Omega_j){\vec d}_j~.
\label{eq:alg-d-inv}
\end{equation}

Let us assume that ${\bf q}_{j}$ have finite range, i.e. 
\begin{equation}
{\bf q}_{j}\equiv 0\qquad |j|>N~,\label{fin-rang}
\end{equation}
for some fixed positive integer $N$. Define
\begin{equation}
{\vec d}_{left}:=
{\vec d}_{j}\qquad j< -N~, \quad  {\vec d}_{right}:=
{\vec d}_{j}\qquad j>N~,\label{eq:dlr-def}
\end{equation}
which are constants. By introducing the $4\times 4$ matrices 
\begin{equation}
R_{j}:= \left(I-\Omega_{j}\right)\prod_{k=-N}^{j-1}
\left(I-2\Omega_{k}\right)~,\qquad 
R:=\prod_{k=-N}^{N}\left(I-2\Omega_{k}\right)~,\label{eq:R-def}
\end{equation}
we derive the relations
\begin{equation}
{\vec d}_{j}=R_{j}\,{\vec d}_{left}\qquad |j|\le N~, \qquad 
{\vec d}_{right}=R\,{\vec d}_{left}~.\label{eq:d-soln}
\end{equation}By the summation form~(\ref{eq:alg-d}) we obtain the formulas
\begin{equation}
{\vec d}_{left}={\vec a}+\sum_j \Omega_jR_j\, {\vec d}_{left},\qquad
R{\vec d}_{left}={\vec a}-\sum_j\Omega_j R_j\, {\vec d}_{left}~,
\label{eq:d-rels}
\end{equation}
by which we find the relations
\begin{equation}
I-R=2\sum_j \Omega_j R_j~,\qquad {\vec a}=\frac{1}{2}(I+R){\vec d}_{left}~.
\label{eq:d-rels-ii}
\end{equation}

Thus, by~(\ref{eq:d-soln}), ${\vec d}_{left}$ alone
suffices to yield $\vph(x,\omega)$ in~(\ref{eq:vph-soln-discr}). With recourse to~(\ref{eq:d-asymp}),
the incoming wavefunction $\vec\psi_{in}$ furnishes immediately the $2\times 2$ vectors 
${\vec d}_{+,left}$ and ${\vec d}_{-,right}$. By writing
\begin{equation}
R=:\left[\matrix{\tilde{\bf R}_{1}& \tilde{\bf R}_2\cr 
\tilde{\bf R}_3& \tilde{\bf R}_4\cr}\right]~,\label{eq:R-Rij} 
\end{equation}
where $\tilde{\bf R}_{k}$ are $2\times 2$ matrices, and using~(\ref{eq:d-soln}) we find
\begin{equation}
{\vec d}_{-,left}=\tilde{\bf R}_4^{-1}({\vec d}_{-,right}-\tilde{\bf R}_3{\vec d}_{+,left})~.
\label{eq:d-l}
\end{equation}For the sake of simplicity, we assume that $\tilde{\bf R}_4$ is non-singular.
The last relation completes the calculation of the four-vector ${\vec d}_{left}$.
Thus, we arrive at the following statement.
\vskip10pt

\noindent{\bf Proposition II.} {\it Equation~(\ref{eq:vphi-discr}), with} ${\bf q}_j\equiv 0$
{\it for} $|j|> N$, {\it is solved by~(\ref{eq:vph-soln-discr}) where} $\vec a$ {\it given 
by~(\ref{eq:d-rels-ii})},
\begin{equation}
\vph_j={\vec d}_{+,j}\,e^{i\sqrt{\omega}\ x_j}+{\vec d}_{-,j}\,e^{-i\sqrt{\omega}\ x_j}~,
\label{eq:vphi-soln-dj}
\end{equation}
{\it and} ${\vec d}_{\pm,j}$ {\it are given by~(\ref{eq:d-soln});} ${\vec d}_{+,left}$
{\it is determined by~(\ref{eq:d-asymp}) and} ${\vec d}_{-,left}$ {\it is determined by~(\ref{eq:d-l})}.
\vskip10pt

The continuous analog of~(\ref{eq:alg-d}), which pertains to the solution 
of~(\ref{eq:S_Eq-FT}), is the Fredholm-type integral equation
\begin{equation}
{\vec d}(x,\omega)={\vec a}(\omega)-\int\limits_{-\infty}^{+\infty}{\rm d}y\ \sg(x-y)\Omega(y,\omega))\,
{\vec d}(y,\omega)~,\label{eq:d-IE}
\end{equation}
where the $4\times 4$ matrix $\Omega$ is 
\begin{equation}
\Omega(x,\omega)=\frac{1}{4\sqrt{\omega}}\left[\matrix{{\bf q}(x)& w(x){\bf q}(x)\cr
                                                        -w^\ast(x){\bf q}(x)& -{\bf q}(x)\cr}\right]~,
\qquad w(x)=e^{-2i\sqrt{\omega}\ x}~.
\label{eq:Omega-def2}
\end{equation}
Differentiation of~(\ref{eq:d-IE}) leads to the Dirac-type equation 
\begin{equation}
\partial_x {\vec d}(x,\omega)=-2\Omega(x,\omega)\,{\vec d}(x,\omega)~.
\label{eq:d-ode}
\end{equation}

Note that the scheme underlying proposition~II corresponds to solving~(\ref{eq:d-IE})
by iterations. Because of the obvious connection of this scheme to the standard theory
of integral equations~\cite{masujima}, we do not discuss~(\ref{eq:d-IE}) any further in this paper.
Once $\vph(x,\omega)$ is known, $\im({\bf M})$ can be calculated via proposition~I in
section~\ref{subsec:imp}. In the next section, we apply proposition~II to a 
delta-function potential~\cite{wu02,margetismyers}.

The procedure of this section, which applies to the Schr\"odinger equation~(\ref{eq:S_Eq})
with a local interaction, can be extended to nonlocal interactions, equation~(\ref{eq:S_Eq-nlc}),
but the algebra is more elaborate. In the next section we consider the case where
the kernel of the interaction becomes a suitable pseudo-potential~\cite{wu02,margetismyers}.

\section{Point interaction}
\label{sec:point-inter}

In this section we revisit the theory of~\cite{margetismyers} in the context
of the present formalism, particularly of
propositions~I,~II in sections~\ref{subsec:imp},~\ref{sec:soln}.
For point interactions and a pure initial state of the memory, 
we calculate the impurity measure and show that the scattering amounts to rotations
of quaternions in the frequency domain. We describe how a class of incoming finite-energy pulses
can produce small impurity.

\subsection{Even wavefunctions}
\label{subsec:even}

First, we consider the Schr\"odinger equation
\begin{equation}
i\partial_{t}\vps(x,t)=-\partial^{2}_{x}{\vec\psi} -
i{\bf q}\delta(x)\vps(0,t)~,\label{eq:delta}
\end{equation}
by which the particle interacts with the memory at the origin. We set
\begin{equation}
{\bf q}=g{\bf u}\qquad  \Vert{\bf u}\Vert =1~,
\end{equation}
i.e., ${\bf u}$ is a unit imaginary quaternion; 
$\Vert {\bf u} \Vert:=\sqrt{\langle {\bf u}\,,\,{\bf u}\rangle}$.
The incoming wavefunction is assumed to be the even part 
of~(\ref{eq:psi-in-f})~\cite{margetismyers}.

By virtue of~(\ref{eq:Deltam-FT}) and~(\ref{eq:prop}), 
the impurity is measured in terms of the quaternionic commutator
\begin{equation}
\Delta{\bf m}=\into\ 
[\widetilde{\bf r}(0,\omega)\,,\,g{\bf u}]\label{eq:Deltam-ev}
\end{equation}
where $\widetilde{\bf r}(0,\omega)$ is 
\begin{equation}
\widetilde{\bf r}(0,\omega)= i\vph(0,\omega)\vph^{\ \dag}(0,\omega)~.
\label{eq:hat-r-ev}
\end{equation}

We now apply the formalism of section~\ref{sec:soln}, in particular proposition~II.
In the present situation we have $N=0$; the associated vector coefficients are  
$\vec{d}_{-1}$, $\vec{d}_{0}$ and $\vec{d}_{1}$. By~(\ref{eq:R-def}) there is only
one propagation matrix, i.e.,
\begin{equation}
R_0= I-\Omega_{0}=\left[\matrix{
{\bf 1}-{g\over 4\sqrt{\omega}}{\bf u}&-{g\over 4\sqrt{\omega}}{\bf u}\cr
{g\over 4\sqrt{\omega}}{\bf u}&{\bf 1}+{g\over 4\sqrt{\omega}}{\bf u}\cr}
\right]~.\label{eq:R0-ev}
\end{equation}
Thus, 
\begin{equation}
{\vec d}_{left}=\vec{d}_{-1}=(I+\Omega_{0})\vec{d}_{0}~,\qquad 
{\vec d}_{right}=\vec{d}_{1}=(I-\Omega_{0})\vec{d}_{0}~.
\label{eq:dpm1-ev}
\end{equation}

The introduction of the $2\times 1$ vector ${\vec d}_{s}$ by
\begin{equation}
\vec{d}_{0}=:\left[\matrix{{\vec d}_{s}\cr {\vec d}_{s}\cr}\right]
\label{eq:ds-def}
\end{equation}
converts~(\ref{eq:dpm1-ev}) to 
\begin{equation}
\vec{d}_{j}=\left[\matrix{\vec{d}_{s}\cr \vec{d}_{s}\cr}\right]-\sg(j)
\frac{g}{2\sqrt{\omega}}{\bf u}\cdot \left[\matrix{\vec{d}_s\cr -\vec{d}_s\cr}\right]~,\qquad j=-1,\,0,\,1~.
\label{eq:dj-ex-ev}
\end{equation}
Vectors relevant to the $\Delta{\bf m}$ of~(\ref{eq:Deltam-ev}) are
\begin{equation}
\vec{d}_{in}:=\vec{d}_{+,left}=\biggl({\bf 1}+\frac{g}{2\sqrt{\omega}}\,{\bf u}\biggr)\vec{d}_s~,
\qquad \vph(0,\omega)=2\vec{d}_{s}~.
\label{eq:s-inout-ev}
\end{equation}

By~(\ref{eq:Deltam-ev}), the entanglement quaternion $\Delta{\bf m}$ reads
\begin{equation}
\Delta{\bf m}=4\into\ [i{\vec d}_s{\vec d}_s^{\ \dag}\,,g{\bf u}]=
\into\ \frac{4g}{p^2}\,e^{-\theta{\bf u}}
[i{\vec d}_{in}{\vec d}_{in}^{\ \dag}\,,\,{\bf u}]e^{\theta{\bf u}}
\label{eq:Deltam-evII}
\end{equation} 
where we conveniently defined the quaternion
\begin{equation}
{\bf p}(\omega):=
{\bf 1}+\frac{g}{2\sqrt{\omega}}\,{\bf u}=:p\,e^{\theta{\bf u}}~.\label{eq:p-def}
\end{equation}
We apply the convention that the magnitude and phase are
\begin{equation}
p(\omega)=
\sqrt{1+\frac{g^2}{4\omega}}~,\qquad 
\theta(\omega)=\arctan\biggl(\frac{g}{2\sqrt{\omega}}\biggr)~.
\label{eq:p-theta-def}
\end{equation}

We now simplify~(\ref{eq:Deltam-evII}) by observing that
${\rm d}\omega/{\rm d}\theta=-(4/g)\,p^2\omega^{3/2}$ and
\begin{equation}
i{\vec d}_{in}{\vec d}_{in}^{\ \dag}=\vert f(\omega)\vert^{2}\,{\bf t}_{in}~,\qquad
{\bf t}_{in}:= i\big(\vec{s}_{in}\vec{s}^{\ \dag}_{in}\big)~,\label{eq:tin-def}
\end{equation}
where $\vec s_{in}$ is introduced in~(\ref{eq:psi-in-f}).
Furthermore, we apply the identity
\begin{equation}
\frac{{\rm d}}{{\rm d}\theta}\big(e^{-\theta{\bf u}}{\bf t}e^{\theta{\bf u}}\big)
=e^{-\theta{\bf u}}[{\bf t}\, ,\, {\bf u}]e^{\theta{\bf u}}~,\label{eq:rot-uth}
\end{equation}
where the operation
$e^{-\theta{\bf u}}{\bf t}e^{\theta{\bf u}}$
is a rotation which leaves the plane spanned by
$\{\sigma_{0},{\bf u}\}$ invariant. We can find two imaginary quaternions $\{{\bf v},{\bf w}\}$ that
are orthogonal to ${\bf u}$; then, let $P_{\bf u}$ be the projection onto the space
spanned by $\{{\bf v}, {\bf w}\}$.  
Equations~(\ref{eq:Deltam-evII})--(\ref{eq:rot-uth}) entail
\begin{equation}
\Delta{\bf m}=-16\into\, \omega^{3/2}\vert f(\omega)\vert^{2}\ 
\frac{{\rm d}}{{\rm d}\omega}\left\{e^{-\theta{\bf u}}\big(P_{\bf u}{\bf t}_{in}\big)e^{\theta{\bf u}}\right\}~.
\label{eq:Deltam-evIII}
\end{equation}
Integration by parts yields
\begin{equation}
\Delta{\bf m}=16\into\ 
e^{-\theta(\omega){\bf u}}\big(P_{\bf u}{\bf t}_{in}\big)e^{\theta(\omega){\bf u}}
\big(\omega^{3/2}\vert f(\omega)\vert^{2}\big)^{\prime}~,\label{eq:Deltam-evIV} 
\end{equation}
where the prime denotes differentiation with respect to $\omega$.
\vskip10pt

\noindent{\bf Example of small-impurity pulse.}
We proceed to describe how a class of incoming finite-energy pulses can
produce an arbitrarily small impurity. Such pulses have of course a narrow spectrum sufficiently
localized at a single frequency.

A simple case of pulses with finite energy is described by 
\begin{equation}
(2\pi)^{-1}\,\omega^{3/2}\vert f(\omega)\vert^{2}=H(\omega -\omega_{0})-H(\omega -K\omega_{0})~,\quad K>1,\ \omega_0>0~,\label{eq:pulse-def}
\end{equation}
where $H$ is the Heavyside function ($H^\prime(\omega)=\delta(\omega)$)
and $(\omega_{0},K)$ are given parameters. 
The amplitude $|f(\omega)|^2$ is scaled by $4\int_0^{+\infty}{\rm d}\omega\,|f|^2\sqrt{\omega}$ so that
the total pulse energy is fixed to unity. This normalization will be carried
out in the impurity measure $\im({\bf M})$ below. In view of~(\ref{eq:Deltam-evIV}), we compute
\begin{equation}
\Delta{\bf m} =16\big(e^{-\theta_{1}{\bf u}}P_{\bf u}{\bf t}_{in}
e^{\theta_{1}{\bf u}}
-e^{-\theta_{2}{\bf u}}P_{\bf u}{\bf t}_{in}e^{\theta_{2}{\bf u}}\big)~,\label{eq:Deltam-ev-ex}
\end{equation}
where 
\begin{equation}
\theta_{1}=\arctan\biggl(\frac{g}{2\sqrt{\omega_{0}}}\biggr)~,\quad
\theta_{2}=\arctan\biggl(\frac{g}{2\sqrt{K\omega_{0}}}\biggr)~;\quad 0<\theta_2<\theta_1<\frac{\pi}{2}~.\label{eq:theta12}
\end{equation}
Note that the operation
$e^{-\theta{\bf u}}P_{\bf u}{\bf t}_{in}e^{\theta{\bf u}}$
rotates $P_{\bf u}{\bf t}_{in}$ by $-2\theta$.

With regard to ${\bf m}_{in}$, by~(\ref{eq:min-pure}) we compute
\begin{equation}
{\bf m}_{in}=4\left(\into\,
\vert f(\omega)\vert^{2}\sqrt{\omega}\right)\, {\bf t}_{in} =
4\ln(K)\,{\bf t}_{in}~.\label{eq:min-ev-ex}
\end{equation}
We notice that $\Delta{\bf m}$ lies in the plane spanned by $\{{\bf v},{\bf w}\}$ 
which is orthogonal to $\{ {\bf 1},{\bf u}\}$. Substituting in~(\ref{eq:prop})
and normalizing properly we find that the impurity measure equals
\begin{equation}
\im({\bf M})=4\sqrt{\frac{\bigl| 4\{1-\cos(2\theta_{1}-2\theta_{2})\}+
\ln(K)\{\cos(2\theta_{1})-\cos(2\theta_{2})\}\bigr|}{(\ln K)^2}}~.
\label{eq:zero-imp}
\end{equation}
The right-hand side of this expression vanishes only for $K=1$,
but becomes arbitrarily small if $K-1\ll 1$. In this limit,
\begin{equation}
\im({\bf M})\sim \sqrt{\frac{2}{3}}\ \frac{g/(2\sqrt{\omega_0})}{\omega_0+g^2/4}\,(\Delta\omega)~,
\quad \Delta\omega:=|K-1|\omega_0~.
\label{eq:imp-ev-smK}
\end{equation}

The behavior $\im({\bf M})=O(\Delta\omega)$ as $\Delta\omega\to 0$ is expected to be generic for any incoming
pulse wavefunction that has spectrum sufficiently localized at $\omega_0$ with
support (bandwidth) of size $\Delta\omega$. The precise prefactor that enters the
formula for $\im({\bf M})$ depends on the specifics of the pulse spectrum.

\subsection{Odd wavefunctions}
\label{subsec:odd}

Next, we turn our attention to the equation~\cite{wu02,margetismyers}
\begin{equation}
i\partial_{t}\vps =-\partial^{2}_{x}\vps +i{\bf q}\delta^{\prime}_{p}(x)
\left(\int\limits_{-\infty}^{+\infty} {\rm d}y\
\delta^{\prime}_{p}(y)\vps(y,t)\right)~,
\label{eq:S_Eq-odd}
\end{equation}
where $\delta_p'(x)$ denotes $\delta'(x)$ modified to remove any
discontinuity at $x=0$ from the function on which it acts~\cite{wu02}:
$\delta_p'(x)g(x):=\delta'(x)[1-\lim_{x\to 0^+}]g(x)$ for $x>0$
and $\delta_p'(x)g(x):=\delta'(x)[1-\lim_{x\to 0^-}]g(x)$ for $x<0$.
For simplicity we write
\begin{equation}
\vps_{x}(0,t) =\int\limits_{-\infty}^{+\infty}{\rm d}y\ \delta^{\prime}_{p}(y)\,\vps(y,t)~.
\label{eq:vpsi-x}
\end{equation}
The incoming wavefunction is assumed to be the odd part 
in~(\ref{eq:psi-in-f})~\cite{margetismyers}. The Fourier transform in $t$ of~(\ref{eq:S_Eq-odd}) gives
\begin{equation}
\omega\vph=-\partial_x^2\vph-i{\bf q}\delta_p^\prime(x)\vph_x(0,\omega)~.
\label{eq:vph-odd}
\end{equation}

By proposition~I in section~\ref{subsec:imp} and equation~(\ref{eq:Deltam}), the entanglement
quaternion reads
\begin{equation}
\Delta{\bf m} =\int\limits_{-\infty}^{+\infty}{\rm d}t\ 
[{\bf n}(0,t)\,,\,{\bf q}]~, 
\label{eq:Deltam-od}
\end{equation}
where 
\begin{equation}
{\bf n}(0,t):= i\,\vps_{x}(0,t)\vps^{\ \dag}_{x}(0,t)~.\label{eq:n-def}
\end{equation}
By use of the Fourier transform of $\vps(x,t)$, we have
\begin{equation}
\Delta{\bf m} =\into\,
[\widetilde{\bf n}(0,\omega)\,,\,{\bf q}]~,\quad \widetilde{\bf n}(0,\omega)=
i\,\vph_x(0,\omega)\vph_x^{\ \dag}(0,\omega)~.\label{eq:Deltam-odII}
\end{equation}

The solution of~(\ref{eq:vph-odd}) reads 
\begin{equation}
\vph(x,\omega)=\vec{a}_{+}(\omega)e^{i\sqrt{\omega}\,x}+
\vec{a}_{-}(\omega)e^{-i\sqrt{\omega}\,x} +
i{\bf q}G_x(x,\omega)\vph_x(0,\omega)~,\label{eq:vph-soln-od}
\end{equation}
where $G_x$ is the derivative of the Green's function~(\ref{eq:G-def}), i.e.
\begin{equation}
G_x(x,\omega)=-{1\over 4}\,\sg(x)\left(e^{i\sqrt{\omega}\,x}+e^{-i\sqrt{\omega}\,x}\right)~.
\label{eq:Gpr-def}
\end{equation}
Note that the pseudo-potential $\delta_p^{\prime}$ gives zero when it acts on $G_x$~\cite{wu02}.
Thus, applying $\delta^{\prime}_{p}$ on~(\ref{eq:vph-soln-od}) yields
\begin{equation}
\vph_x(0,\omega)=i\sqrt{\omega}
\big\{\vec{a}_{+}(\omega)-\vec{a}_{-}(\omega)\big)\}~. 
\label{eq:vphi-x-od}
\end{equation}
Substitution of~(\ref{eq:vphi-x-od}) into~(\ref{eq:vph-soln-od}) with 
$\vec{d}=(\vec{d}_{+},\,\vec{d}_{-})^{T}$ leads to the formula
\begin{equation}
\vec{d}(x,\omega)=\left[\matrix{{\bf 1}+\frac{g}{4}\sg(x)\sqrt{\omega}\,{\bf u}
&-\frac{g}{4}\sg(x)\sqrt{\omega}\,{\bf u}\cr
\frac{g}{4}\,\sg(x)\sqrt{\omega}\,{\bf u}& {\bf 1}-\frac{g}{4}\,\sg(x)\sqrt{\omega}\,{\bf u}\cr}
\right]\left[\matrix{\vec{a}_{+}\cr \vec{a}_{-}\cr}\right]~.\label{eq:d-od}
\end{equation}

According to the imposed antisymmetry, we introduce the $2\times 1$ vector $\vec{d}_a$ in
\begin{equation}
\left[\matrix{\vec{a}_{+}\cr \vec{a}_{-}\cr}\right]=
\left[\matrix{\vec{d}_a\cr -\vec{d}_a\cr}\right]~.
\label{eq:a-od}
\end{equation}Thus, we have
\begin{eqnarray}
\vec{d}_{+}(x,\omega)&=&\biggl\{{\bf 1}+\frac{g}{2}\sg(x)\sqrt{\omega}\,{\bf u}\biggr\}\vec{d}_{a}~,\nonumber\\
\vec{d}_{-}(x,\omega)&=&-\biggl\{{\bf 1}-\frac{g}{2}\sg(x)\sqrt{\omega}\,{\bf u}\biggr\}\vec{d}_{a}~.
\label{eq:dpm-od}
\end{eqnarray}
Furthermore, by~(\ref{eq:vph-soln-od}),
\begin{equation}
\vph_x(0,\omega)=2i\sqrt{\omega}\,\vec{d}_{a}~.
\label{eq:vph0-od}
\end{equation}

The vector $\vec{d}_{in}$ corresponding to in-states can be 
$\vec{d}_{-}(+\infty,\omega)$ or $\vec{d}_{+}(-\infty,\omega)$; compare to~(\ref{eq:d-asymp}).
For example,
\begin{equation}
\vec{d}_{in}(\omega)=\biggl({\bf 1}-\frac{g}{2}\sqrt{\omega}\,{\bf u}\biggr)\vec{d}_{a}~.
\label{eq:din-od}
\end{equation}
Hence, by analogy with section~\ref{subsec:even} it makes sense to define
\begin{equation}
{\bf 1}+\frac{g}{2}\sqrt{\omega}\,{\bf u} =: p(\omega)e^{\theta{\bf u}}~,\label{eq:pth-od}
\end{equation}
where 
\begin{equation}
p(\omega)=\sqrt{1+\omega g^{2}/4}~,\quad \theta(\omega)=
\arctan\biggl({g\over 2}\sqrt{\omega}\biggr)~. \label{eq:pth-odII}
\end{equation}

Equation~(\ref{eq:Deltam-odII}) for $\Delta{\bf m}$ becomes
\begin{equation}
\Delta{\bf m}=\into\ \omega\,\frac{4g}{p^2}\, 
e^{\theta\,{\bf u}}[i\vec{d}_{in}\vec{d}_{in}^{\ \dag}\,,\,{\bf u}]e^{-\theta{\bf u}}~.
\label{eq:Deltam-odIII}
\end{equation}
The assumption of a pure initial state amounts to using $i\vec{d}_{in}\vec{d}_{in}^{\ \dag}$
from~(\ref{eq:tin-def}), i.e. $i\vec{d}_{in}\vec{d}_{in}^{\ \dag}=\vert f(\omega)\vert^2\,{\bf t}_{in}$
where ${\bf t}_{in}=i\,\vec{s}_{in}\vec{s}_{in}^{\ \dag}$ and $\vec{s}_{in}$
is introduced in~(\ref{eq:inc}).
Note that
\begin{equation}
\frac{{\rm d}\omega}{{\rm d}\theta}=\frac{4}{g}\sqrt{\omega}\,p^2.~\label{eq:deriv-od}
\end{equation}
Thus, by analogy with the symmetric case (section~\ref{subsec:even}) we find the formula
\begin{equation}
\Delta{\bf m}=16\into\
\biggl\{e^{\theta(\omega){\bf u}}
\big(P_{\bf u}{\bf t}_{in}\big)e^{-\theta(\omega){\bf u}}\biggr\}\,
\big(\omega^{3/2}\vert f(\omega)\vert^{2} \big)^{\prime}~.
\label{eq:Deltam-odIV}
\end{equation}
The impurity $\im({\bf M})$ follows by the procedure of section~\ref{subsec:even}.

A simple example of an incoming pulse wavefunction is described again by
$(2\pi)^{-1}\omega^{3/2}\,\vert f(\omega)\vert^2=H(\omega-\omega_0)-H(\omega-K\omega_0)$.
The analysis for the impurity follows the steps of section~\ref{subsec:even} and is omitted here.
Equation~(\ref{eq:zero-imp}) should be recovered, where the angles $\theta_1$ and $\theta_2$ are now defined by
\begin{equation}
\theta_1=\arctan\biggl(\frac{g\sqrt{\omega_0}}{2}\biggr)~,\quad
\theta_2=\arctan\biggl(\frac{g\sqrt{\omega_0 K}}{2}\biggr)~;\ 0<\theta_1 <\theta_2<\frac{\pi}{2}~.
\end{equation}

\section{Conclusion}
\label{sec:conclusion}

We introduced a general formulation of the nonrelativistic scattering
from a two-state quantum memory in one space dimension. The key feature is to view
the interaction potential as an imaginary quaternion. In the case with point interactions,
scattering from the memory amounts to a rotation in the
frequency domain of an appropriately defined incoming quaternionic state.

We described the time evolution of the (entanglement) reduced density matrix in
terms of the space integral of appropriate quaternionic commutators. By identifying
quaternions with four-vectors we point out that, because of the
space integration, the quaternions involved in the entanglement evolution are time-like. 
Accordingly, the impurity measure for the final memory state is
described by a time-integral containing the Minkowski norm of time-like, hyperbolic quaternions.
In the special case of narrow-band pulse wavefunctions, the resulting impurity $\im({\bf M})$
is generically of the order of the pulse bandwidth.

This work can be useful for addressing several questions. It is tempting to study successive scatterings
from a quantum memory modeled by point interactions. In this case incoming states may not
be pure but incoming quaternions are successively rotated
in an appropriate sense in the frequency domain. An interesting question is how the impurity changes by
this process. It is expected that $\im({\bf M})$ always increases in such a case,
especially if $\im({\bf M})$ is thought of as `entropy' in a sense~\cite{wooters}.
There is no rigorous justification of this claim at the moment.
The connection of $\im({\bf M})$ to interference effects critical to quantum computing such
as those discussed in~\cite{jaeger} was not addressed by our analysis.

Another possible extension is the case of a $n$-state memory. A related issue
is to define the appropriate algebra of $n\times n$ matrices that
describe scattering in this context.

Finally, it is interesting to consider relativistic massive particles within
the present framework. A starting point would be the case of particles with spin $1/2$.
A perhaps naive model problem is the one-dimensional scattering from 
a two-state memory in the setting of Dirac's equation.
In this case, the particle-memory system is described by a $8\times 1$ vector field.
The study of this process by use of an analogous formalism is the subject of future work.

\section*{Acknowledgments}
\label{sec:ack}
We thank John M Myers for invaluable discussions.

\appendix

\section{Proof of proposition~I}
\label{app}

In this appendix we prove proposition~I of section~\ref{subsec:imp}.
In particular, we show that an arbitrary null quaternion can be written
as a tensor product of the form $\vps\vps^{\ \dag}$. Furthermore, we show that
the impurity measure $\im({\bf M})$ is given by~(\ref{eq:prop}).

Suppose we have a null quaternion
\begin{equation}
{\bf r}=ir^{0}\sigma_{0}+\sum_{a=1}^3r^{a}\bs_{a}~,\label{eq:null-r}
\end{equation}
and let us write $\vec{r}:=(r^{1},r^{2},r^{3})$ for the space vector. The `nullity'
property means that
$(r^{0})^{2}=\vert\vec{r}\vert^{2}$; we normalize so that $r^0=1/2$. 
The nullity is invariant under scaling. So, 
if we set ${\bf h}=2{\bf r}$ we have that ${\bf h}$ is also null with $h^{0}=1$ and 
$\vert\vec{h}\vert =1$. An appropriate stereographic projection can identify 
the unit vector $\vec{h}$ with the complex number
\begin{equation}
z:=\frac{h^{1}-ih^{2}}{1-h^{3}}~.\label{eq:z}
\end{equation}
The inversion of this mapping yields 
\begin{eqnarray}
r^{1}&=&\frac{(z+z^{\ast})/2}{1+zz^{\ast}}~,\nonumber\\
r^{2}&=&-\frac{(z-z^{\ast})/2i}{1+zz^{\ast}}~,\nonumber\\
r^{3}&=&\frac{(zz^{\ast}-1)/2}{1+zz^{\ast}}~.
\label{eq:r}
\end{eqnarray}

The substitution $z:=z_{1}/z_{2}$ in~(\ref{eq:r}) where $\vert z_1\vert^2+\vert z_2\vert^2=1$ gives
\begin{eqnarray}
r^{0}&=&\big(z_{1}z^{\ast}_{1}+z_{2}z^{\ast}_{2}\big)/2~,\nonumber\\
r^{1}&=&\big(z_{1}z^{\ast}_{2}+z^{\ast}_{1}z_{2}\big)/2~,\nonumber\\
r^{2}&=&\big(z_{1}^{\ast}z_{2}-z_{1}z_{2}^{\ast}\big)/2i~,\nonumber\\
r^{3}&=&\big(z_{1}z^{\ast}_{1}-z_{2}z^{\ast}_{2}\big)/2~.
\label{eq:r-z12}
\end{eqnarray}
These relations show that an arbitrary null quaternion can be written as the tensor product
$\vps\vps^{\ \dag}$ where
\begin{equation}
\vps:=\left[\matrix{z_{1}\cr z_{2}\cr}\right]~,\quad \vert z_1\vert^2+\vert z_2\vert^2=1~.
\label{eq:z12}
\end{equation}

Next, we show formula~(\ref{eq:prop}) for $\im({\bf M})$.
By setting ${\bf M}=-i{\bf r}$, it is straightforward to calculate 
\begin{equation}
{\bf r}^{2}=\big(-(r^{0})^{2}-\vert\vec{r}\vert^{2}\big)\sigma_{0}
+2ir^{0}\sum_ar^{a}\bs_{a}~.\label{eq:r2}
\end{equation}
By invoking the algebra of the Pauli matrices, we find 
\begin{equation}
1-\tr\big({\bf M}^{2}\big) =2\big[(r^{0})^{2}-\vert\vec{r}\vert^{2}\big] =
-2\big({\bf r}{\bf r}^{d}\big)~,\label{eq:tr}
\end{equation}
so that 
\begin{equation}
\im\big({\bf M}\big) =\sqrt{2}\sqrt{\vert{\bf r}{\bf r}^{d}\vert}=
\sqrt{2}\sqrt{\vert\big<{\bf r},{\bf r}\big>\vert}=\sqrt{2}\Vert {\bf r}\Vert~,\label{eq:im-norm}
\end{equation}
which confirms~(\ref{eq:prop}) of proposition~I in section~\ref{subsec:imp}.
In the above, ${\bf r}=(r^{0},r^{1},r^{2},r^{3})$ is a four-component vector and $<\cdot\,,\,\cdot>$ 
is the Minkowski inner product.

\end{document}